# Safety benefit assessment of autonomous emergency braking and steering systems for the protection of cyclists and pedestrians based on a combination of computer simulation and real-world test results

Jordanka Kovaceva[1], András Bálint[1], Ron Schindler[1], Anja Schneider[2]

[1]Department of Mechanics and Maritime Sciences, Chalmers University of Technology

[2]AUDI AG

**Abstract**

Cyclists and pedestrians account for a significant share of fatalities and serious injuries in the road transport system. In order to protect them, advanced driver assistance systems are being developed and introduced to the market, including autonomous emergency braking and steering systems (AEBSS) that autonomously perform braking or an evasive manoeuvre by steering in case of a pending collision, in order to avoid the collision or mitigate its severity.

This study proposes a new prospective framework for quantifying safety benefit of AEBSS for the protection of cyclists and pedestrians in terms of saved lives and reduction in the number of people suffering serious injuries. The core of the framework is a novel application of Bayesian inference in such a way that prior information from counterfactual simulation is updated with new observations from real-world testing of a prototype AEBSS.

As an illustration of the method, the framework is applied for safety benefit assessment of the AEBSS developed in the European Union (EU) project PROSPECT. In this application of the framework, counterfactual simulation results based on the German In-Depth Accident Study Pre-Crash Matrix (GIDAS-PCM) data were combined with results from real-world tests on proving grounds.

The proposed framework gives a systematic way for the combination of results from different sources and can be considered for understanding the real-world benefit of new AEBSS. Additionally, the Bayesian modelling approach used in this paper has a great potential to be used in a wide range of other research studies.

# 1  Introduction

Cyclists and pedestrians account for a significant share of fatalities and serious injuries in the road transport system. According to the Annual Accident Report by the European Commission (EC, 2018), 8% of road fatalities in the European Union (EU) were cyclists and 22% were pedestrians; consequently, these two groups comprise 30% of the 25,651 fatalities in the EU in 2016. In order to reduce the number of crashes between passenger cars and vulnerable road users (VRUs) such as pedestrians and cyclists, advanced driver assistance systems (ADAS) are being developed and introduced to the market. A specific type of ADAS called autonomous emergency braking and steering systems (AEBSS) are those that detect a pending collision and autonomously perform braking or an evasive manoeuvre by steering to avoid the collision or mitigate their severity. These systems are expected to reach the market in the near future and the development and assessment of these ADAS is important to quantify the expected real-world benefit of these systems.

Several approaches have been proposed for assessing the expected real-world safety benefit of ADAS (Carter, Brugett, Srinivasan, & Ranganathan, 2009; Page, et al., 2015; Sander, 2018). In general, a distinction is made between *retrospective* and *prospective* safety benefit assessment. *Retrospective assessment* is based on observed real-life data after the systems are implemented in the vehicles. Retrospective assessment has been performed e.g. by analysis of insurance claims data (Doyle, Edwards, & Avery, 2015; Isaksson-Hellman & Lindman, 2016; Cicchino, 2017; Cicchino, 2018; Kuehn, Hummel, & Bende, 2009), by meta-data analysis in Fildes et al. (2015), by using national crash databases as in Sternlund et al. (2017), Gårder & Davies (2006) or Persaud et al. (2001) and by analysing naturalistic driving (ND) data (McLaughlin, Hankey, & Dingus, 2008; van Noort, Faber, & Bakri, 2012). The assessment provides the true representation of the effect of the systems, but it may require a long time until the systems can be evaluated.

*The prospective approach* performs the assessment before the implementation of the actual systems in the vehicles, for example, by real-world testing, driving simulator studies, or computer simulations. Real-world testing means physical testing in a controlled environment and is often used to determine if the system works according to specifications (Nilsson, 2014; Edwards, et al., 2015). Real-world testing has the advantage of testing the actual system in a safe environment which ensures high fidelity; however, the number of tests is usually limited due to the cost of performing the tests, and the interactions are with dummies and without a driver. The cost is decreased in driving simulator studies in which human drivers are interacting with the model of the system being evaluated, see Aust, Engström, & Viström (2013). Further cost reduction can be obtained in computer simulations, where all components (driver, vehicle, environment) are being modelled. Computer simulations can be divided in simulations of artificial scenarios, see Wang et al. (2016), Yanagisawa et al. (2017) and Jeong & Oh (2017), and counterfactual simulations ("what-if simulations"). In counterfactual simulation, a re-analysis of real-world crashes under various assumptions is performed (Bärgman, Boda, & Dozza, 2017). The counterfactual simulation approach enables performing a high number of simulations and can be applied in the early stage of development of the system; however, idealization of the system models is common, and the reliability of the results depends on the validation and verification of the models.

After the data has been collected by, e.g. the real-world testing or counterfactual simulations, statistical methods are used to quantify the safety benefit of ADAS in terms of saved lives and avoided injuries. Table 1 below summarizes results from a literature review of relevant assessment methods, the underlying data sources and whether and how results from different sources were combined in the assessment.

**Table 1: Data sources and data combination methods in the research literature addressing safety benefit assessment of ADAS.**

| Reference | Assessment type | Data sources and combination methods |
|---|---|---|
| Carter, Brugett, Srinivasan, & Ranganathan (2009) | Prospective | Virtual simulations (proposal for harmonization) |
| Page, et al. (2015) | Prospective | Virtual simulations (proposal for harmonization) |
| Sander (2018) | Prospective | Counterfactual simulations |
| Doyle, Edwards, & Avery (2015) | Retrospective | Insurance database |
| Isaksson-Hellman & Lindman (2016) | Retrospective | Insurance database |
| Cicchino (2017) | Retrospective | Crash database analysis |
| Cicchino (2018) | Retrospective | Crash database analysis |
| Fildes, et al. (2015) | Retrospective | Multinational crash data |
| Sternlund, et al. (2017) | Retrospective | Crash database analysis |
| Kuehn, Hummel, & Bende (2009) | Retrospective | Insurance database |
| McLaughlin, Hankey, & Dingus (2008) | Retrospective | ND collected during real crashes and near-crashes |
| van Noort, Faber, & Bakri (2012) | Retrospective | ND from field operational test |
| Bayly, et al. (2007) | Retrospective and prospective | Literature review but no combination of results |
| Jeong & Oh (2017) | Prospective | Microscopic traffic simulator |
| Lindman, et al. (2010) | Prospective | Counterfactual simulations |
| Edwards, et al. (2015) | Retrospective | Real-world testing |
| Wang, et al. (2016) | Prospective | Virtual simulations |
| Kusano & Gabler (2012) | Prospective | Simple counterfactual simulations |
| Fahrenkrog, et al. (2019) | Prospective | Virtual simulations |
| Lee, et al. (2019) | Retrospective | Real-world testing |
| Zhao, Ito, & Mizuno (2019) | Prospective | Counterfactual simulations on ND crash data |
| Haus, Sherony, & Gabler (2019) | Prospective | Counterfactual simulations |
| Wimmer, et al. (2019) | Prospective | Virtual simulations (review for harmonization) |
| Gårder & Davies (2006) | Retrospective | Crash database analysis |
| Persaud, et al. (2001) | Retrospective | Crash database analysis |
| Gårder, Leden, & Pulkkinen (1998) | Prospective | Quantitative expert judgment model, combined in a Bayesian approach |
| Hauer (1983) | Prospective | Quantitative expert judgment model, combined in a Bayesian approach |

As shown in Table 1, safety benefit assessment is typically based on a single data source and there is no commonly accepted way for the combination of results from different sources. For example, Bayly et al. (2007) analysed both retrospective and prospective benefit estimations but report all results independently of each other and do not provide a method to combine the results from both fields. Notable exceptions from this trend are those studies in which Bayesian statistical methods have been used to accumulate evidence from different sources about the effect of countermeasures, starting with two seminal papers by Hauer (1983) and including Gårder, Leden, & Pulkkinen (1998) reviewed in Table 1 above. Our paper follows this line of research to provide a systematic approach for combining simulation results and real-world test results.

*The aim of this paper is to propose a new prospective framework for safety benefit assessment of AEBSS for the protection of cyclists and pedestrians based on a systematic combination of simulation results and real-world test results.*

Using Bayesian methods for the combination of results from different sources in the assessment is motivated by a theorem that if mathematical representations of the prior information and the sampling model represent a rational person's beliefs, Bayesian inference is a mathematically optimal way of updating prior information with new observations (Hoff, 2009). Bayesian methods have been applied in the context of traffic safety research in the above mentioned references Hauer (1983) and Gårder, Leden, & Pulkkinen (1998) as well as several other studies (Miaou & Lord, 2003; Mitra & Washington, 2007; Huang & Abdel-Aty, 2010; Xie, Dong, Wong, Huang, & Xu, 2018; Morando, 2019). However, the idea of defining a prior based on simulation results and updating it with real-world test results of higher fidelity is new and can be applied and extended in various ways.

The framework is exemplified on a new generation of AEBSS developed in the EU project Proactive Safety for Pedestrians and Cyclists (PROSPECT), where the systems perform autonomous emergency braking and, in longitudinal scenarios, steering (Aparicio, et al., 2017). The safety benefits of the PROSPECT AEBSS in terms of saved lives and avoided serious injuries were estimated in the project report by Kovaceva et al. (2018) using the current framework, and the corresponding socio-economic benefit in the year 2030 was quantified under various assumptions. This paper gives a more focused description of the safety benefit framework excluding the quantification of the benefit in monetary terms, and a sensitivity analysis is conducted to examine whether the findings are consistent when parameters are varied.

This paper is structured as follows. Section 2 describes the safety benefit assessment framework and its elements. Section 3 exemplifies the method on the PROSPECT AEBSS and the corresponding results are presented in Section 4, including a sensitivity analysis. The developed framework, its application in PROSPECT and potential future work are discussed in detail in Section 5. Finally, the main conclusions are summarized in Section 6.

## 2   The assessment framework

The proposed safety benefit assessment framework is illustrated in Figure 1 below. In the definition of the framework, it is assumed that a prototype of the AEBSS whose safety benefit needs to be assessed is available for testing and that a computer model representing the main features of the AEBSS can be defined. Additionally, we also assume the availability of real-world crash data (or naturalistic driving data) that can be used to identify *use cases (UCs),* i.e. a set of relevant real-world crash scenarios that the AEBSS is intended to address. Counterfactual simulation and prototype testing of the AEBSS in the use cases yield two separate sets of results. These results are combined in a systematic way in a Bayesian statistical approach, with simulation results representing prior information regarding the effectiveness of the AEBSS which is then updated with the test results to obtain a posterior benefit representing information from both sets of results. The posterior benefit is extrapolated to quantify the maximum potential benefit of AEBSS in a larger region of interest ("target region"), which can then be adjusted for the expected market penetration and user acceptance to estimate the safety benefit, e.g. in terms of the number of lives saved and reduction in the number of people suffering serious injuries.

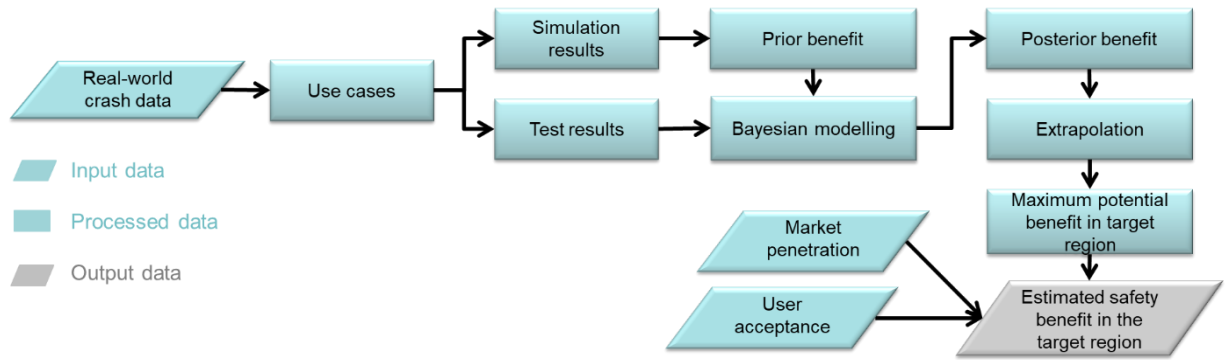

**Figure 1: The proposed framework for safety benefit assessment of an AEBSS.**

The elements in Figure 1 are discussed further in the following sections.

## 2.1    Data

The assessment framework expects two types of input data on crashes. The first type of data can be extracted from in-depth crash data as in the studies by Lindman et al. (2010), Eichberger et al. (2010), Kusano and Gabler (2012), and Rosén (2013) or naturalistic driving data as in Bärgman et al. (2015; 2017). These data are used to derive use cases. The second type of crash data used in the assessment framework is official road crash data from a larger target region. Such data typically does not contain detailed crash information in terms of kinematic parameters and detailed injury information and would therefore be insufficient in itself for the AEBSS assessment. Consequently, information from both data sources are needed to quantify the effect of the AEBSS in the target region, by extrapolation as described in Section 2.5.

## 2.2    Simulation results

The proposed framework requires simulation of the relevant UCs with and without the assessed AEBSS to quantify its effect. In the application of the framework, virtual counterfactual simulations using relevant models for vehicle dynamics, sensors, and AEBSS algorithms are performed on each of the crashes. Here existing simulation tools can be used (Wang, et al., 2016; Sander, 2018) and additional work may be required to include a relevant representation of the AEBSS in the simulation tool with sufficient validity and reliability.

## 2.3    Test results

Beside simulation results, the proposed framework uses a possibly smaller sample of high-fidelity results for an accurate and effective assessment. This can be generally real-world prototype testing, which includes the actual system in the vehicle but is expensive to perform and may therefore yield a relatively smaller set of results.

## 2.4 Bayesian modelling

A novel element of the benefit assessment framework in Figure 1 is a Bayesian approach for combining simulation results and test results. In this model, prior information from the simulation results is updated with new observations from the test results. The theoretical foundations and applications of Bayesian methods are described e.g. in Kruschke (2015) or in Hoff (2009) whose terminology is used throughout this paper. The output of the Bayesian modelling is the posterior benefit, which is an estimate of the maximum safety benefit of the AEBSS in the database used for generating the UCs, based on both simulation results and test results.

With the implementation of the AEBSS, some crashes may be avoided, while others may be mitigated or unchanged. Consequently, for each UC, two separate models are considered: one estimates the probability of the collision avoidance by the safety system (see Section 2.4.1) and the other model estimates the collision speed, given that the collision is not avoided (Section 2.4.2). These two models are used in a stochastic approach to estimate the posterior benefit, as described in Section 2.4.3.

### 2.4.1 Modelling the probability of crash avoidance

For each UC, the initial speeds of cars in the simulations are matched to those in the real-world tests. For each initial speed prescribed by the matching, the prior probability of crash avoidance is assumed to follow a Beta($a, b$) distribution whose parameters are determined by the number of crashes avoided (denoted by $a$) and the number of crashes not avoided (denoted by $b$), respectively. As the Beta distribution is a conjugate prior for binomial sampling (see Section 3.1.1 of Hoff (2009)), the posterior distribution is a Beta($c, d$) with $c = a + w \sum y_i$ and $d = b + wn - w \sum y_i$, where $n$ is the number of new observations, $w$ is a parameter in the model representing the weight of test results compared to simulation results, and for $i \in \{1, 2, \ldots, n\}$, $y_i$ takes value 1 if the crash is avoided in test $i$ and 0 otherwise.

For the considered initial speeds, the posterior estimate for the crash avoidance probability is taken as the median of the Beta($c, d$) distribution and a 90% confidence interval is obtained by taking the 0.05 and 0.095-quantiles. The estimates and confidence intervals are extended to all initial speeds by fitting a logistic curve to the values obtained as above.

### 2.4.2 Modelling collision speed in case of a crash

Modelling of collision speed in case of a crash based on simulation results can be considered as a standard statistical modelling task and various models (e.g. regression models) can be constructed for this purpose. An example for such modelling is described in Section 3.4. This model can then be used for determining a prior distribution required by the Bayesian modelling framework.

The modelling of collision speed in case of a crash in this paper is focused on the case that while collisions with the AEBSS included in the vehicle may occur in re-simulation of crashes, all collisions are avoided in real-world testing. This assumption is discussed in Section 5.3 where it is also argued that this case is highly relevant in real-world applications. Under this assumption, there are no new observations for the model estimating collision speed when the crash is not avoided from real-world testing, hence the posterior distribution equals the prior and is fully determined by the simulation results.

### 2.4.3 Estimation of the posterior benefit

The estimation of the posterior safety benefit is done using a variant of the dose-response model (Korner, 1989; Kullgren, 2008; Bálint, Fagerlind, & Kullgren, 2013). This model estimates the number of people suffering injuries of a given type or severity based on crash frequency and injury risk, with respect to a crash severity parameter which in this paper is selected to be the collision speed.

For any given UC and a specific speed value $v$ measured in kilometres per hour, crash frequency at $v$ is defined as $f(v)$ = the number of car-to-VRU crashes within the UC occurring at collision speed $v$, and for a fixed injury severity (e.g. seriously injured), the injury risk $r(v)$ is the risk of sustaining an injury of the given severity. Furthermore, we define $L$ as the greatest value $v$ such that $f(v) > 0$ (i.e. a speed value which is the upper limit of collision speeds within the UC). The dose-response model estimates the number of VRUs sustaining an injury of the given severity within the UC, denoted by $E(N)$, as follows:

$$E(N) = \int_0^L f(v) r(v) \, dv \tag{1}$$

In this formula, the dependence of these quantities on the UC is suppressed in the notation for simplicity. As AEBSS can potentially avoid a crash or change the collision speeds for those crashes that cannot be avoided, for their assessment, $f(v)$ needs to be replaced by a new crash frequency function $f_{new}(v)$; the details of how to estimate this function are described below, see equation (2). Assuming that this function is known, an estimate corresponding to $E(N)$ can be computed using the same injury risk function $r(v)$ by replacing the original crash frequency function $f(v)$ in (1) by $f_{new}(v)$.

The original crash frequency function per UC can be approximated based on the collisions speeds in the crashes used in the simulation (Section 2.2) and the injury risk function can be taken from the research literature (if available) or constructed from the crash data. The way the crash frequency curve is transformed by the AEBSS can be estimated based on the results described in Sections 2.4.1 and 2.4.2, as follows. For each simulated crash $c$, the model developed in Section 2.4.1 specifies a probability $p(c)$ of the crash being avoided, and an estimate $\tilde{v}_{coll}(c)$ regarding the collision speed in case the crash is not avoided (which happens with probability $1 - p(c)$), rounded to the closest integer value. Therefore, for a given UC, the transformed crash frequency function can be estimated by defining, for each nonnegative integer value $v$, the quantity

$$f_{new}(v) = \sum_{c \in UC} (1 - p(c)) \, 1_{\{\tilde{v}_{coll}(c) = v\}} \tag{2}$$

where $c \in UC$ means that the crash is included in the given UC and $1_{[\tilde{v}_{coll}(c) = v]}$ is an indicator function taking value 1 if the estimated collision speed for the crash rounded to the closest integer equals the specified value $v$ and 0 otherwise.

Having specified all functions as above, the dose-response model quantifies the expected number of cases with the considered injuries (e.g. fatalities and serious injuries) per UC with and without the assessed AEBSS. Those can then be summarized to quantify the posterior benefit for all UCs, and using the confidence intervals specified in the models in Sections 2.4.1 and 2.4.2 yields confidence intervals for the reductions as well. The model can also be used to consider a subset of crashes within the UC which may be necessary input to the method extrapolating the local benefit to EU level, as described in Section 2.5.

## 2.5 Extrapolation

In-depth crash data that is required to perform the previous steps is typically collected in small regions, but the effectiveness of the AEBSS should preferably be quantified in larger regions beyond the sampling area. Therefore, the results from the posterior benefit need to be extrapolated to a target region. This can be performed by a decision tree method (Broughton, et al., 2010; Ferreira, 2015; Kreiss, et al., 2015), as follows. First, a decision tree for in-depth data can be built, and then a decision tree with the same rules can be calculated for the target region data. The frequency of the injuries in the terminal nodes for both trees can be used to calculate the extrapolation factors according to

$$Factor_{i,n} = \frac{(Frequency\ in\ target\ region)_{i,n}}{(Frequency\ in\ the\ in\text{-}depth\ data)_{i,n}} \qquad (3)$$

where $i$ is injury level and $n$ is terminal node number. The output of this step is the maximum potential benefit of the assessed AEBSS in the target region expressed as reduction of casualties (i.e. fatalities or the number of seriously injured people).

### 2.6   Estimated safety benefit in the target region

In order to estimate the expected casualty reduction from the maximum potential values resulting from the extrapolation step, user acceptance and market penetration need to be taken into consideration. In this context, market penetration is the percentage of vehicles in traffic that include the safety system. When a safety system is introduced into the market, it will not directly be available in all vehicles, but its installation rate will likely increase over the years (see e.g. Schneider (2016) and HLDI (2017)). User acceptance of the system is defined here as the conditional probability that the system is turned on, given that it is available in a vehicle (e.g. if the system cannot be turned off, then this is always 100%). This parameter is relevant since even if systems are available in all vehicles but poorly designed, users will refrain from using them, resulting in a situation as if the system did not exist.

In this paper, a linear relationship is assumed between these parameters and safety benefit, i.e. the safety benefit is estimated to be the product of the maximum potential benefit of the AEBSS in the target region with the market penetration $0 < p < 1$ and user acceptance $0 < u < 1$ values.

### 2.7   Sensitivity analysis

The weight parameter $w$ is freely chosen in Section 2.4.1 and the most relevant value for this parameter may vary depending on the simulation and test setup. Therefore, to understand the dependence of the results on this parameter, a sensitivity analysis is recommended at each application of the framework. An example analysis is described in Section 3.7.

## 3   Application of the framework in PROSPECT

The application of the framework is demonstrated in the context of the AEBSS developed in the PROSPECT project. In this example, we focus mainly on cyclist UCs and less on pedestrian UCs.

### 3.1   Crash data from GIDAS and CARE

This study used pre-crash kinematics data as input to the framework, from crashes between passenger cars and VRUs from the Pre-Crash Matrices (PCM) of the German In-Depth Accident Study (GIDAS) database. Twelve UCs were derived from the crash configurations, 9 between cars and cyclists and 3 between cars and pedestrians (see Figure 2): UC1 (vehicle turns left, oncoming cyclist), UC2 (vehicle turns right and cyclist in same direction), UC3 (crossing situation with cyclist from the right), UC4 (crossing situation with cyclist from the left), UC5 (vehicle turns right, cyclist crosses from the right), UC6 (vehicle turns left, cyclist crosses from the left), UC7 (cyclist passes a car, while driver opens the door), UC8 (cyclist passes a car, while passenger opens the door), UC9 (car rear-ends cyclist travelling in the same direction), UC10 (pedestrian crosses from the right), UC11 (pedestrian crosses from the right, direct vision obscured), UC12 (car passes pedestrian in same direction). For the details of selecting these use cases, see Wisch et al. (2017) and Kovaceva et al. (2018).

In GIDAS crash data from the years 1999-2015 was used, which included 4,406 crashes between passenger cars and cyclists out of which 4,231 were crashes with only 2 participants (one car and one cyclist). From these cases, 77% (3,239 from 4,231) are covered by the accident types classified in the car-cyclist UCs. The total number of crashes between passenger cars and pedestrians in GIDAS is 1,768, and 1,574 crashes are between one car and one pedestrian. The accident types classified in the car-pedestrian UCs cover 44% of these cases (687 from 1,574). The PCM data was filtered to identify input crashes with reliable information on pre-crash trajectories of the vehicles, to be used in the counterfactual simulations. The resulting dataset, used in the rest of the paper, contains 1,943 crashes matched to the 12 UCs, shown in Figure 2.

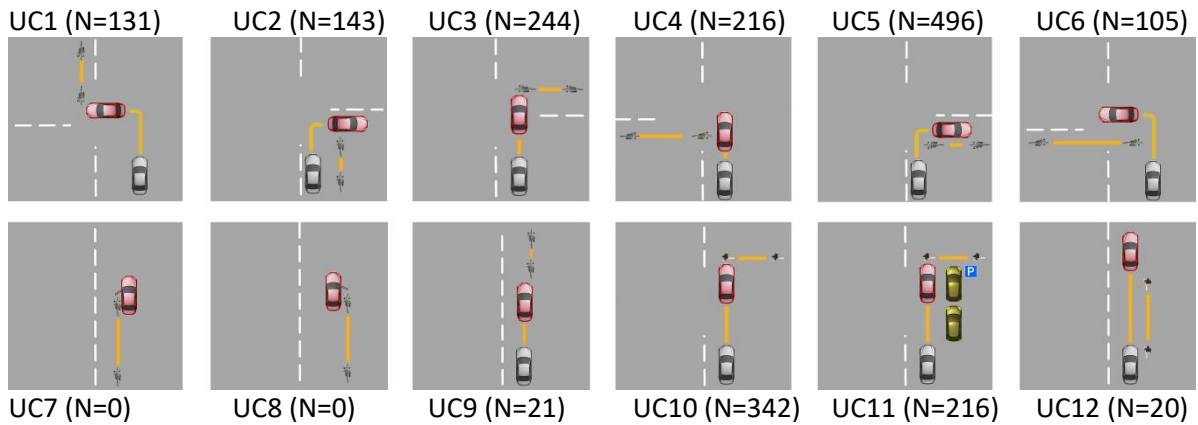

**Figure 2: All cyclist and pedestrian UCs and number of crashes per UC.**

Additional to the definition of UCs, GIDAS was used to construct injury risk curves estimating the probability of the VRU suffering injuries of different severity at various collision speeds, separately for each UC and separately for cyclists and pedestrians. An ordered probit model (Liddell & Kruschke, 2018) was used to construct the injury risk curves, see Appendix 9.1 for the details.

The second type of crash data required for the method is official road accident data on a European level, contained in the Community Database on Accidents on the Roads in Europe (CARE). Data from CARE was queried for car-to-cyclist crashes and car-to-pedestrian crashes in EU-28 for the years 2009-2013 with additional selection requirements described in Wisch, et al. (2016). The proportion of data that corresponds to the cyclist UCs and pedestrian UCs amounts to 307,643 car-to-cyclist crashes and 199,693 car-to-pedestrian crashes. This data was used to quantify the effect of the safety systems in Europe by extrapolation, see Sections 2.5 and 3.5.

### 3.2    Simulation – description of algorithms

In order to simulate the changes of the crash caused by the defined systems, the rateEFFECT tool was used, see Wille et al. (2012). Four system algorithms that represent models of the prototype AEBSS were implemented: Algorithms 1, 2 and 3 performed only emergency braking for avoiding the crash while Algorithm 4 performed both emergency braking and steering in the longitudinal use cases (UC9 and UC12).  In each algorithm, low-level comfort braking with a deceleration level of 4 m/s$^2$ was initiated based on moderate values of the TTC between 1.5 and 0.5 seconds. Algorithm 1 initiated emergency braking for smaller values of TTC.  Algorithm 2 performed emergency braking based on the TTC and if the collision is unavoidable by evasive manoeuvres of both the vehicle and the cyclist. Algorithm 3 initiated emergency braking based on the TTC and if the collision is unavoidable by evasive manoeuvres from the vehicle, not considering evasive manoeuvres of the cyclist. Algorithm 4 initiated emergency steering if the collision is unavoidable by braking, and if the collision is unavoidable by both steering and braking, then a mitigating brake manoeuvre is initiated to reduce collision speed as much as possible. In each algorithm, emergency braking corresponded to a deceleration level of 9 m/s$^2$. Further details of the algorithms can be found in Krebs, Kunert, Arbitmann, & Othmezouri (2018).

## 3.3 Testing in PROSPECT

Vehicle-based testing activities on closed test tracks were performed to evaluate the performance of the developed AEBSS prototypes according to a test protocol, aligned with current consumer testing procedures. All tests were conducted with driving robots, including a steering robot as well as brake and acceleration actuation. This allowed the vehicle dynamics to be controlled over the whole test run and ensured high repeatability of each test. The VRUs were represented with a dummy of a bicyclist or a pedestrian. The velocity of the bicyclist and pedestrian dummy was set to 15 km/h and 5 km/h, respectively. The output of testing with the implemented prototype systems used for the safety benefit assessment was whether a crash was avoided, and in case of a collision, the speed reduction of the tested vehicle.

## 3.4 Bayesian modelling

Modelling the probability of the collision avoidance based on the simulation results and matched test results is performed as described in Section 2.4.1, with Table 3 in Section 4.2 describing the matching between test results and use cases. A parameter value of $w = 2$ was used in this application, assigning double weight to test results compared to simulation results, and the effect of other values of $w$ on the results were explored in the sensitivity analysis (see Section 3.7). Extending the posterior estimate for the crash avoidance probability from the test speeds to all initial speeds was done by fitting a logistic curve as described in Section 2.4.1 for Algorithms 1 to 3, but for Algorithm 4, the logistic curve did not give an appropriate fit and a polynomial curve was used instead.

For each UC and each algorithm as implemented in the simulation, a linear regression model is constructed to model the collision speed when the crash is not avoided, considering the following variables: initial speed of the car; initial speed of the VRU; longitudinal distance; lateral distance; sight obstruction (No/Not permanent/Permanent/Other); location (Urban/Rural). All subsets of these variables were considered as covariates and the linear regression model with the lowest Akaike Information Criterion (AIC) value was selected as the final model; see Akaike (1974) for more details on AIC. However, in UC9 and UC12, the models with the lowest AIC value showed multi-collinearity of the covariates and hence the models with the second lowest AIC value were selected.

Finally, following the notation in Section 2.4.3, the original crash frequency function $f(v)$ was derived from the PCM data described at the beginning of Section 3 for each UC and was combined with injury risk functions $r(v)$ that were constructed for fatal, serious and slight injuries of cyclists and pedestrians based on GIDAS data, see Section 9.1 in the Appendix. This allowed the estimation of the posterior benefit per UC.

## 3.5 Extrapolation to EU level

To quantify the benefit in the target region EU-28, extrapolation was performed from in-depth data in GIDAS to European data in CARE using the decision tree method described in Section 2.5. The variables used for the extrapolation and the resulting factors are described in Section 4.4.

## 3.6 Estimated safety benefit in the EU

An illustration of the possible values for user acceptance and market penetration was employed. From studies within the PROSPECT project a user acceptance of 82% was assumed at market introduction, see Kovaceva et al. (2018); therefore, this value is used as an example in this paper. For market penetration, an example value of 20% is considered; for airbags, it took 10 years to reach this market penetration and 15 years to reach 50% (Schneider, 2016; HLDI, 2017).

## 3.7 Sensitivity analysis

As indicated in Section 2.7, it is investigated how results concerning the estimated casualty reduction of the AEBSS change by changing the value of the weight parameter $w$. Additionally, in this paper introducing the assessment framework, sensitivity of results with respect to other methodological choices and decisions have also been investigated. The following three aspects have been addressed by the sensitivity analysis:

a) *Injury risk functions.* As described in Section 3.1, the injury risk functions used in this paper were constructed using ordinal probit regression. In this part, it is investigated whether and how the results change if the injury risk functions are derived using logistic regression instead.
b) *Frequentist approach.* Instead of using a systematic Bayesian approach to combine test results with simulation results as specified in Section 3.4, it is analysed how the results are affected if test results are added to the sample of simulation results with a multiplicity determined by the weight parameter $w$, thus circumventing the update step described in Section 2.4.1.
c) *Different values of $w$.* The parameter $w$ determines the weight of test results compared to simulation results, and all results in Sections 4.1 to 4.5 as well as those addressing point b) above are derived with $w = 2$. Assuming higher fidelity of real-world tests compared to simulation results suggests that $w \geq 1$ should be taken but does not in itself determine the optimal value of $w$. As the value of $w$ can be freely changed in the method, various values are considered in this part of the sensitivity analysis, including $w = 0$ (i.e. test results are disregarded), $w = 1$ (test results are considered equally relevant as simulation results) and $w = 10$ (assigning substantially greater weight to test results compared to simulation results).

The results of the sensitivity analysis are summarized in Section 4.6.

## 4 Results for the PROSPECT AEBSS

The final results of the application of the safety benefit assessment framework are specified for all four algorithms used to represent the PROSPECT AEBSS in counterfactual simulations (see Section 3.2), while results of the intermediate steps are presented for Algorithm 1 only.

## 4.1 Simulation results from rateEFFECT

The summary of the simulation results from the rateEFFECT tool for Algorithm 1 per UC are shown in Table 2. In each simulation, the collision was either avoided or mitigated (i.e. the collision speed was reduced with the implemented system). The results for the other algorithms are given in Appendix 9.2. The percentage of avoided crashes is greater than mitigated crashes in UC1, UC4, UC5, UC6, UC9 for all three algorithms addressing these UCs. For UC2, UC10, UC11 and UC12, the percentage of mitigated crashes is larger than the avoided crashes.

Table 2: Frequency of total, avoided and mitigated crashes for Algorithm 1 per UC.

| Use Case | Total | Algorithm 1 | | | |
|---|---|---|---|---|---|
| | | Avoided | Mitigated | % Avoided | % Mitigated |
| UC1 | 131 | 112 | 19 | 85.5 | 14.5 |
| UC2 | 143 | 53 | 90 | 37.1 | 62.9 |
| UC3 | 244 | 124 | 120 | 50.8 | 49.2 |
| UC4 | 216 | 142 | 74 | 65.7 | 34.3 |
| UC5 | 496 | 409 | 87 | 82.5 | 17.5 |
| UC6 | 105 | 83 | 22 | 79.0 | 21.0 |
| UC9 | 21 | 16 | 5 | 76.2 | 23.8 |
| UC10 | 342 | 156 | 186 | 45.6 | 54.4 |
| UC11 | 216 | 48 | 168 | 22.2 | 77.8 |
| UC12 | 20 | 5 | 15 | 25.0 | 75.0 |

## 4.2 Test results in PROSPECT

In vehicle-based testing of the prototype systems in PROSPECT, all collisions were avoided. Table 3 below summarizes the number of tests by initial speed for each use case.

**Table 3: Number of tests by use case and car initial speed used to update simulation results. The tests for UC9 and UC12, listed in the bottom rows, included steering while all other tests included braking only.**

| Use case | Car initial speed (km/h) | | | | | | | TOTAL |
|---|---|---|---|---|---|---|---|---|
| | 10 | 15 | 20 | 30 | 40 | 50 | 60 | |
| UC1 | 1 | 1 | 1 | | | | | 3 |
| UC2 | 3 | 3 | | | | | | 6 |
| UC3 | | | 1 | 1 | 1 | 1 | | 4 |
| UC4 | | | | 3 | 3 | 7 | | 13 |
| UC5 | 1 | 1 | 1 | | | | | 3 |
| UC6 | 1 | 1 | 1 | | | | | 3 |
| UC10 | | | 1 | 1 | 1 | 1 | | 4 |
| UC9 | | | | 1 | 1 | 1 | 1 | 4 |
| UC12 | | | | 1 | 1 | 1 | 1 | 4 |

Note that UC7 and UC8 are not listed as they are not addressed by the assessment method. Additionally, there were no test results matched to UC11. The tests for UC9 and UC12 included steering while all other tests included braking only. Therefore, the test results for UC9 and UC12 are used to update the simulation results for Algorithm 4 (braking and steering algorithm) for the corresponding UCs while the other test results are used to update simulation results for Algorithms 1-3.

## 4.3 Prior and posterior models for the PROSPECT AEBSS

As described in Section 2.4.1, the prior probability of crash avoidance was estimated by UC and by car initial speed, and this was updated using the corresponding test results. The result for UC6 and car initial speed of 15 km/h is illustrated in Figure 3 below.

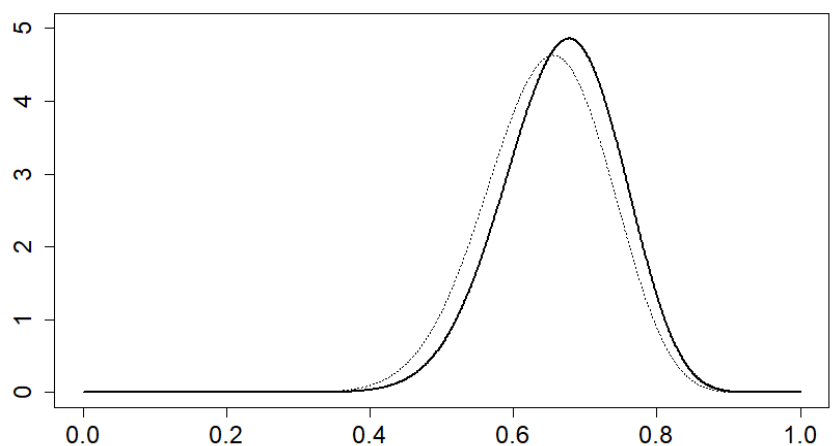

Figure 3: Bayesian update of the crash avoidance probability for UC6 and car initial speed 15 km/h. The dotted line shows the density function of the prior distribution, which is Beta(20,11), and the solid line is the density function of the posterior distribution, which is Beta(22,11).

The curve fitting procedure on the points determined by the posterior distribution as described in Section 2.4.1 yielded logistic models for the posterior probability of crash avoidance. The coefficients of these models are specified for Algorithm 1 in Table 4 and for Algorithm 2 and 3 in Appendix 9.2.

Table 4: Posterior models for the probability of crash avoidance for Algorithm 1.

|  | Algorithm 1 | |
| --- | --- | --- |
|  | Coefficient of car initial speed $\beta_1$ | Intercept $\beta_0$ |
| UC1 | -0.205 | 5.774 |
| UC2 | 0.031 | -0.882 |
| UC3 | -0.071 | 2.004 |
| UC4 | -0.051 | 2.383 |
| UC5 | -0.016 | 1.730 |
| UC6 | 0.033 | 1.171 |
| UC9 | -0.915 | 51.054 |
| UC10 | -0.088 | 2.457 |
| UC11 | -0.033 | -0.288 |
| UC12 | -0.370 | 14.146 |

The parameters in Table 4 determine the posterior probability of crash avoidance for all initial speeds. For example, for UC1 and Algorithm 1, the posterior probability of crash avoidance based on the logistic model is estimated as $p = e^{\beta_0 + \beta_1 v_{init}}/(1 + e^{\beta_0 + \beta_1 v_{init}})$, where $\beta_0 = 5.774$ is the intercept, $v_{init}$ is the initial speed of the car and $\beta_1 = -0.205$ is the corresponding coefficient.

Based on the models of crash avoidance probability and collision speed in case of a crash modelled by linear regression, the posterior benefits regarding the reduction of fatalities and the number of seriously injured were estimated with a dose-response analysis as described in Section 2.4.3. The results for each UC are shown in Figure 4.

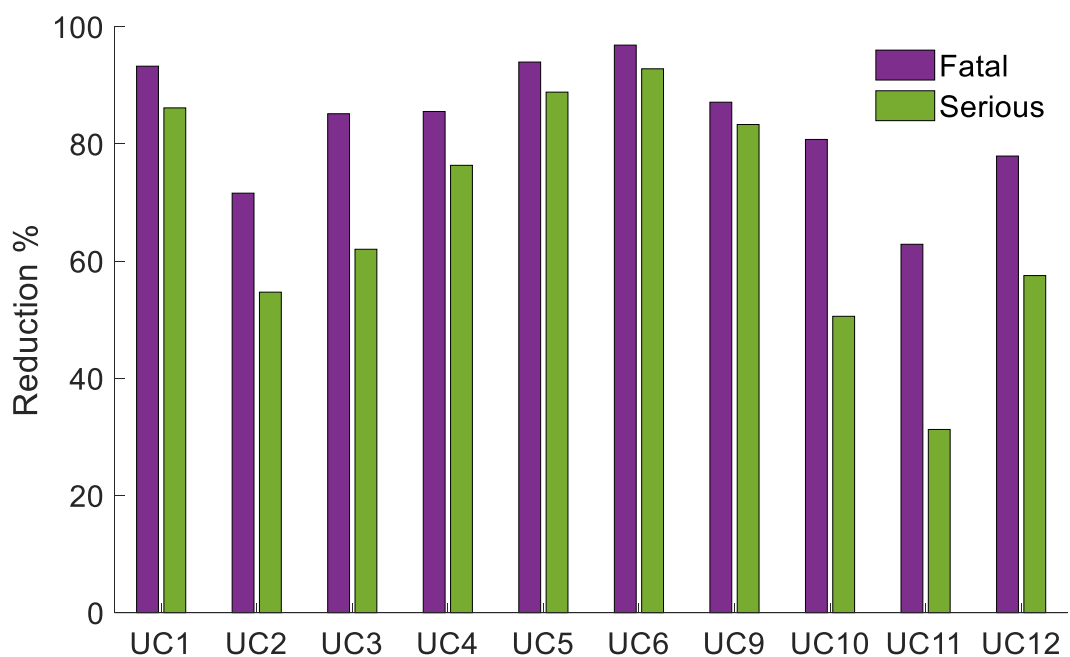

Figure 4: Posterior benefit for Algorithm 1 per UC: percent of injury reduction for cyclists (UC1-UC9) and pedestrians (UC10-UC12).

For the cyclist UCs shown in Figure 4, the maximum benefit is achieved for UC6 (vehicle turning left and cyclist coming from left), while the minimum benefit is for UC2 (vehicle turning right and cyclist coming from the same direction). For the pedestrian UCs (i.e. UC10-UC12 in Figure 4), casualty reduction is greatest for UC12 (car passes pedestrian in same direction) and smallest for UC11 (pedestrian crosses from the right, direct vision obscured).

### 4.4 Extrapolation to EU level

As indicated in Section 3.5, a decision tree is built based on GIDAS data for the classification of the cyclist injuries in the car-to-cyclist crashes, using a set of relevant variables (e.g. weather, surface, light, site, gender and age). The same classification criteria are then applied to CARE data. The corresponding decision trees are provided in the Appendix 9.3. A comparison of the classification tree results from both databases yield weighting factors that are used for the extrapolation of the benefit to EU level, see Table 5. For example, the first factor in the first row can be interpreted so that one cyclist that is fatally injured and younger than 55 years from GIDAS corresponds to 522 cyclists on EU level.

**Table 5: Factors for extrapolation of cyclist and pedestrian injuries from GIDAS to CARE.**

|  |  | Fatal | Serious |
|---|---|---|---|
| **Cyclist** | Age <= 55 | 522 | 160 |
|  | Age > 55 and Not Urban | 948 | 301 |
|  | Age > 55 and Urban | 652 | 144 |
| **Pedestrian** | Age > 55 | 425 | 228 |
|  | Age <= 55 and Not Daylight | 408 | 199 |
|  | Age <= 55 and Daylight | 767 | 120 |

The reduction of cyclist injuries from GIDAS (Section 3.1) is extrapolated to CARE by using the extrapolation factors for cyclists in Table 5 leading to an estimate of the maximal annual reduction of cyclist casualties due to the PROSPECT AEBSS on EU level, shown in Table 6.

**Table 6: Maximum potential reduction of cyclist injuries for one year by different algorithms for UC1 to UC9 on EU level. The estimate is given together with the lower and higher bound interval (in brackets). The "braking only" algorithms (Algorithms 1-3) address all cyclist use cases while the steering and braking algorithm addresses UC9 only.**

|  | UC1 to UC9 | | | UC9 only |
|---|---|---|---|---|
|  | Algorithm 1 – braking | Algorithm 2 – braking | Algorithm 3 – braking | Algorithm 4 – steering and braking |
| **Fatal** | 693 (650-719) | 666 (617-698) | 673 (625-702) | 102 |
| **Serious** | 7,762 (6,353-8,613) | 7,322 (6,045-8,225) | 7,435 (6,020-8,323) | 294 |

The reduction of the number of injured pedestrians from GIDAS is extrapolated to CARE by using the extrapolation factors for pedestrians in Table 5 estimating of the maximal annual reduction of pedestrian casualties on EU level, see Table 7.

**Table 7: Maximum potential reduction of pedestrian injuries for one year by different algorithms for UC10 to UC12. The estimate is given together with the lower and higher bound interval (in brackets). The "braking only" algorithms (Algorithms 1-3) address all pedestrian use cases while the steering and braking algorithm addresses UC12 only.**

|  | UC10 to UC12 | | | UC12 only |
|---|---|---|---|---|
|  | Algorithm 1 – braking | Algorithm 2 – braking | Algorithm 3 – braking | Algorithm 4 – steering and braking |
| **Fatal** | 1,147 (1,070-1,214) | 1,067 (991-1,139) | 1,074 (995-1,145) | 78 (59-82) |
| **Serious** | 4,543 (3,454-5,605) | 3,859 (2,909-4,858) | 3,939 (2,964-4,947) | 487 (364-515) |

The above results can be put in context by noting that there were 2,064 cyclist fatalities and 5,527 pedestrian fatalities in the EU in 2016 according to the Annual Accident Report (EC, 2018) and results in Wisch et al. (2016) indicate that 49% of cyclist fatalities and 61% of pedestrian fatalities result from crashes involving one passenger car and one cyclist, respectively pedestrian. Consequently, and taking the lower and upper bounds in Table 6 and Table 7 into account, the PROSPECT AEBSS with the "braking only" function could potentially reduce car-to-cyclist fatalities by 61%-71% and car-to-pedestrian fatalities by 29%-36%, depending on the algorithm used for the representation of the PROSPECT AEBSS. Additionally, if the steering-and-braking function is considered in the longitudinal use cases (i.e. UC9 and UC12) instead of the braking only algorithm, the number of cyclist fatalities decreases with additional 6-24 cases and seriously injured with 31-91 cases, while pedestrian fatalities decrease with additional 1-6 cases and the number of seriously injured with 115-212 cases; see Table 15 and Table 16 in Appendix 9.2.

As indicated, Table 6 and Table 7 provide estimates of the maximum potential benefit of the assessed safety system, which could be achieved if all cars in EU-28 were equipped with the system and it would be functioning at all times in each car. The next section adjusts these values for more realistic conditions regarding market penetration and user acceptance.

## 4.5 Estimated safety benefit in the EU

Taking the assumed increasing market penetration and user acceptance into account described in Section 3.6 amounts to multiplying the values in Table 6 and Table 7 by $up = 0.82 * 0.2 = 0.164$, estimating the safety benefit on EU level separately for cyclists and pedestrians and depending on the algorithm used for the representation of the PROSPECT AEBSS. Taking the minimum values of the lower bounds across algorithms 1-3 gives a lower estimate of the benefit with respect to the "braking only" algorithms, giving a reduction of at least 264 fatalities and 1464 serious injured for cyclists and pedestrians together. Taking the maximum values of the higher bounds gives the corresponding higher bounds of 317 fatalities and 2332 seriously injured. Furthermore, if the steering-and-braking function is applied in UC9 and UC12, that gives a further reduction of 1-5 fatalities and 24-50 seriously injured. Considering the statistics on the number of cyclist and pedestrian fatalities in the EU specified in Section 4.4 above, the results indicate that the PROSPECT AEBSS at the specified market penetration and user acceptance levels could avoid 5-6% of all cyclist fatalities and 10-12% of car-to-cyclist fatalities, respectively 3-4% of all pedestrian fatalities and 5-6% of car-to-pedestrian fatalities in the EU.

## 4.6 Sensitivity analysis

In this section, it is reviewed how the results related to the posterior benefit (before the extrapolation step) are affected by methodological decisions and parameter values. As there were no large differences between the results for the different "braking only" algorithms (Algorithms 1 to 3) and the results for braking and steering (Algorithm 4) are based on a small sample, only the results for Algorithm 1 are provided in this section. For easier comparison, the values obtained with the proposed method presented in the earlier sections (see Figure 4) are provided below as "reference values".

### 4.6.1 Injury risk function and statistical approach

The posterior benefit estimates corresponding to the application of the framework with injury risk functions constructed with logistic regression (instead of ordinal probit regression), respectively the results obtained by the frequentist approach (instead of a Bayesian update) described in points a) and b) in Section 3.7 are provided in Table 8 below.

Table 8: The effect of constructing injury risk curves (IRC) from logistic regression, respectively using a frequentist approach instead of a Bayesian update, on casualty reduction (posterior benefit).

| | | IRCs from logistic regression | Reference values | Frequentist approach |
|---|---|---|---|---|
| **FATALITIES** | Total | 83% (76%-89%) | 78% (70%-85%) | 79% (73%-86%) |
| | Total cyclist | 90% (81%-94%) | 86% (77%-92%) | 87% (79%-92%) |
| | Total pedestrian | 81% (74%-87%) | 76% (68%-82%) | 77% (71%-84%) |
| **SERIOUSLY INJURED** | Total | 58% (44%-69%) | 60% (47%-71%) | 63% (54%-71%) |
| | Total cyclist | 75% (59%-85%) | 76% (61%-85%) | 77% (67%-84%) |
| | Total pedestrian | 39% (28%-51%) | 43% (32%-55%) | 48% (41%-58%) |

The results using a different injury risk function are generally similar to the reference values, except that the reduction of fatalities using the injury risk function constructed with logistic regression is somewhat higher; the differences are in the range up to 6% and the largest differences are observed for cyclist fatalities and pedestrian fatalities. However, the reduction for seriously injured pedestrians is somewhat lower in this case. For the frequentist statistical approach, the results are most different from reference values for serious injuries of pedestrians. Generally, the non-Bayesian method gives greater casualty reduction in all cases, but the differences are typically smaller than 5%.

### 4.6.2 Weight of test results compared to simulation results

The results for the posterior benefit with different weights are summarized in Table 9 indicating that the reduction increases with larger weights. This is not surprising considering that collisions were avoided in each test. As described in Section 3.4, the weight parameter $w = 2$ was used to obtain the results shown in Figure 4, hence the corresponding values are the reference values in this table.

Table 9: The effect of changed weights of test results compared to simulation results on casualty reduction (posterior benefit), including the estimated reduction as well as lower and upper bounds.

| | | w=0 | w=1 | w=2 (Reference) | w=10 |
|---|---|---|---|---|---|
| **FATALITIES** | Total | 78% (70%-84%) | 78% (70%-85%) | 78% (70%-85%) | 79% (72%-86%) |
| | Total cyclist | 85% (75%-91%) | 86% (76%-91%) | 86% (77%-92%) | 89% (81%-93%) |
| | Total pedestrian | 76% (68%-82%) | 76% (68%-82%) | 76% (68%-82%) | 76% (69%-83%) |
| **SERIOUSLY INJURED** | Total | 58% (45%-69%) | 59% (46%-70%) | 60% (47%-71%) | 63% (50%-73%) |
| | Total cyclist | 73% (57%-83%) | 75% (59%-84%) | 76% (61%-85%) | 80% (66%-88%) |
| | Total pedestrian | 43% (31%-55%) | 43% (32%-55%) | 43% (32%-55%) | 45% (33%-57%) |

## 5 Discussion

This paper proposes a new prospective framework for assessment of AEBSS for the protection of cyclists and pedestrians and demonstrates its application on the proactive safety systems developed in the PROSPECT project. General aspects of the framework in relation to other benefit assessment methods are discussed in Section 5.1 and lessons learned from the illustrative application of the framework are addressed in Section 5.2. Finally, potential directions for future studies are specified in Section 5.3.

### 5.1 Research methodology

The framework is based on a systematic combination of results from simulation and real-world prototype testing. The advantage of using counterfactual simulations is that they are risk-free, reproducible, time-efficient and allow performing multiple tests. At the same time, the quality of results depends on how well the implemented models represent reality. On the other hand, the prototype real-world testing includes real vehicles and environment but limited number of tests. The relatively low number of tests makes it challenging to understand the transformation of the crash frequency for the full range of crash severity values and to specify the new crash frequency function $f_{new}(v)$ in the dose-response model (see Bálint et al. (2013) for an example). Combining the results from both counterfactual simulations and real-world testing as proposed in this paper allows the utilization of the advantages of each data type and helps overcoming some of the challenges inherent in methods based on one data type only.

The combination of the different types of results in the proposed framework is based on Bayesian statistical modelling. Generally, Bayesian inference allows modelling data with various sorts of statistical distributions and, in contrast to traditional frequentist methods, allows expressing uncertainty in model parameters in form of credibility, by giving access to the full posterior distribution of each parameter (Hoff, 2009; Kruschke, 2015). The derived posterior distributions of the model parameters from a Bayesian model can become the prior distributions for future studies. One example is that the posterior distributions obtained in this study can potentially be used as priors in a retrospective approach once the AEBSS from PROSPECT become available on the market.

Finally, the proposed Bayesian modelling has suitable interfaces required by the initiative P.E.A.R.S. for harmonization of ADAS assessment and therefore can be incorporated into the theoretical framework suggested in Page et al. (2015). Using knowledge synthesis from simulations and tests, it is possible to derive more comprehensive and representative conclusions regarding safety benefit of AEBSS.

### 5.2 Application of the framework and sensitivity analysis

The method was illustrated on AEBSS from the EU project PROSPECT, with the purpose of demonstrating how the framework is used in a real-world situation. For a detailed discussion of the results concerning the PROSPECT AEBSS, see Kovaceva et al. (2018), where it is indicted that the results are subject to various limitations and should be interpreted from this perspective. Notwithstanding, the analysis demonstrates the applicability of the assessment framework and the results show the capabilities of next generation safety systems for VRU protection. In this paper, there is less focus on the numerical values of the results and only those aspects of the illustrative case are discussed in this section that have implications on the safety benefit assessment framework, e.g. the influence of various methodological choices on the results.

In the sensitivity analysis, it was investigated how results concerning the estimated casualty reduction of the AEBSS system change by changing the following three aspects: injury risk functions (IRF), frequentist approach instead of Bayesian modelling and weight parameter $w$. Based on the results in Section 4.6, the method seems relatively robust on the construction method of IRF, at least in the given context. Other procedures for constructing IRFs, e.g. those described in Petitjean et al. (2012) and Yoganandan et al. (2016) could also be tested in future analysis. The frequentist approach overestimated the safety benefit in the application of the method, but the differences were small. These results indicate that reasonable approximations of the proposed framework can be obtained using a simplified, frequentist statistical modelling. The added value of the Bayesian approach as proposed in this paper is that it gives a more detailed characterization of uncertainty as well as a more coherent conceptual framework than frequentist approaches. Changing the weight parameter $w$, determining the relevance of the test results in comparison to the simulation results, had a tangible effect on the reduction estimates for cyclists. Therefore, it is important that the weight parameter is chosen according to the requirements of the analysis in future applications of the method, reflecting the perceived relevance of test results compared to simulation results for the assessment.

### 5.3   Future work

The modelling was done separately by use case, but in order to match simulation results and test results, initial speed of the car was used as a predictor for each use case for the crash avoidance probability model. This was suboptimal for those use cases in which crash avoidance is only loosely related to initial speed, like UC2. In future projects planning to apply the proposed framework, the test protocol should be designed to enable an appropriate matching between the results from different sources. Additionally, modelling of collision speed in case of a crash was done independently from the probability of a collision and was based only on those cases in which the collision was not avoided. A potential refinement of the current method would be a unified model that draws conclusions from all simulated cases, including the information about speed reduction also in those cases where the crash was avoided.

The simulations as described in Section 3.2 focused on autonomous actions of the vehicle and did not include a driver model, which was found to have a large effect on the estimate of safety benefit of ADAS using counterfactual simulations in Bärgman, Boda, & Dozza (2017). Thus, in future research and applications of counterfactual simulations for the safety benefit assessment of AEBSS, algorithms for the system and driver behaviour models should be studied jointly. Additionally, market penetration and user acceptance of the AEBSS may depend on several factors, including marketing strategies of car manufacturers, design of the systems, laws and regulations and consumer testing protocols, which makes it very difficult to obtain accurate predictions. The linear relationship between user acceptance and market penetration towards the reduction of casualties may also be an oversimplification according to Sander (2018), hence modelling the dependence on these parameters could also be refined in future research.

A further limitation regarding the proposed framework is that a single in-depth data source may not capture all relevant aspects of the crash population in the target region. The extrapolation method corrects for some of the differences in terms of the variables used in the process but may not adjust for all differences. In future applications of the proposed framework, this aspect could be improved by using in-depth crash data from several different parts of the target region that may allow the characterization of relevant local differences within the target region.

Finally, test results in PROSPECT were special in the sense that the crash was avoided in each considered real-world test. Consequently, the model for collision speed in case of a crash did not need to be updated with test results. In more general situations, an extension of the current model may need to be applied, including the identification of appropriate prior distributions for the model and an application of Markov chain Monte Carlo methods, see Hoff (2009), to obtain the posterior distribution. Note, however, that the situation observed in PROSPECT is not uncommon: considering the price and vulnerability of VRU dummies used for real-world testing, future research projects may also be motivated to perform only those tests in which collision avoidance can be expected.

## 6   Conclusion

An assessment framework was defined in this paper to prospectively estimate the safety benefit of AEBSS for the protection of cyclists and pedestrians. The core of the assessment framework is a Bayesian update of prior information from simulation with new observations from real-world testing. The framework was applied for safety benefit assessment of the proactive safety systems developed in the PROSPECT project and a sensitivity analysis was conducted showing relatively little sensitivity of results on the main assumptions made in this paper. The presented benefit assessment framework can be used in future studies, and the results have potential implications for policies and regulations in understanding the real-world benefit of new active safety systems. Additionally, the Bayesian modelling approach of defining priors based on initial information of potentially lower fidelity and updating it with results of presumably high fidelity used in this paper has a great potential to be used in other studies.

# 7 References

Akaike, H. (1974). A new look at the statistical model identification. *IEEE Transactions on Automatic Control 19*, 716-723.

Aparicio, A., Sanz, L., Burnett, G., Stoll, H., Arbitmann, M., Kunert, M., . . . Gavrila, D. (2017). Advancing active safety towards the protection of vulnerable road users: the PROSPECT project. *ESV.* Detroit, USA.

Aust, M., Engström, J., & Viström, M. (2013). Effects of forward collision warning and repeated event exposure on emergency braking. *Transportation Research Part F: Traffic Psychology and Behaviour*, 34-46.

Bálint, A., Fagerlind, H., & Kullgren, A. (2013). A test-based method for the assessment of pre-crash warning and braking systems. *Accident Analysis and Prevention, 59*, 192-199.

Bärgman, J., Boda, C., & Dozza, M. (2017). Counterfactual simulations applied to SHRP2 crashes: The effect of driver behavior models on safety benefit estimations of intelligent safety systems. *Accident Analysis and Prevention 102*, 165–180. doi:10.1016/j.aap.2017.03.003

Bärgman, J., Lisovskaja, V., Victor, T., Flannagan, C., & Dozza, M. (2015). How does glance behavior influence crash and injury risk? A 'what-if' counterfactual simulation using crashes and near-crashes from SHRP2. *Transportation Research Part F: Traffic Psychology and Behaviour*.

Bayly, M., Fildes, B., Regan, M., & Young, K. (2007). Review of crash effectiveness of intelligent transport systems. *Emergency*, 3, 14.

Broughton, J., Keigan, M., Yannis, G., Evgenikos, P., Chaziris, A., Papadimitriou, E., . . . J., T. (2010). Estimation of the real number of road casualties in Europe. *Safety Science, 48*(3), 365-371. doi:https://doi.org/10.1016/j.ssci.2009.09.012

Carter, A., Brugett, A., Srinivasan, G., & Ranganathan, R. (2009). SAFETY IMPACT METHODOLOGY (SIM): EVALUATION OF PRE-PRODUCTION SYSTEMS. *ESV*, (pp. 1-14). Stuttgart, Germany.

Cicchino, J. (2017). Effectiveness of forward collision warning and autonomous emergency braking systems in reducing front-to-rear crash rates. *Accident Analysis & Prevention, Volume 99*, 142-152.

Cicchino, J. (2018). Effects of lane departure warning on police-reported crash rates. *Journal of Safety Research, Volume 66*, 61-70.

Doyle, M., Edwards, A., & Avery, M. (2015). AEB REAL WORLD VALIDATION USING UK MOTOR INSURANCE CLAIMS DATA . *ESV.* Gothenburg, Sweden.

EC. (2018). *Annual Accident Report 2018.* Available online at https://ec.europa.eu/transport/road_safety/sites/roadsafety/files/pdf/statistics/dacota/asr2018.pdf; accessed January 31, 2019.

Edwards, M., Nathanson, A., Carroll, J., Wisch, M., Zander, O., & Lubbe, N. (2015). Assessment of Integrated Pedestrian Protection Systems with Autonomous Emergency Braking (AEB) and

Passive Safety Components. *Traffic Injury Prevention*, 2-11. doi:10.1080/15389588.2014.1003154

Eichberger, A., Tomasch, E., Rohm, R., Steffan, H., & Hirschberg, W. (2010). Detailed analysis of the benefit of different traffic safety systems in fatal accidents. . *Proceedings of the annual EVU meeting*, (pp. 301-315). Prague.

Fahrenkrog, F., Wang, L., Platzer, T., Fries, A., Rasich, F., & Kompass, K. (2019). PROSPECTIVE SAFETY EFFECTIVENESS ASSESSMENT OF AUTOMATED DRIVING FUNCTIONS –FROM THE METHODS TO THE RESULTS. *ESV, 2019.* Eindhoven, NL.

Ferreira, S. F. (2015). The quality of the injury severity classification by the police: An important step for a reliable assessment. *Safety Science, 79*, 88-93. doi:https://doi.org/10.1016/j.ssci.2015.05.013

Fildes, B., Keall, M., Bos, N., Lie, A., Page, Y., Pastor, C., . . . Tingvall, C. (2015). Effectiveness of low speed autonomous emergency braking in real-world rear-end crashes. *Accident Analysis & Prevention, Volume 81*, 24-29.

Gårder, P., & Davies, M. (2006). Safety Effect of Continuous Shoulder Rumble Strips on Rural Interstates in Maine. *Transportation research record, 1953(1)*, 156-162.

Gårder, P., Leden, L., & Pulkkinen, U. (1998). Measuring the Safety Effect of RaisedBicycle Crossings Using a New Research Methodology. *Transportation Research Record, 1636(1)*, 64-70.

Hauer, E. (1983). An application of the likelihood/Bayes approach to the estimation of safety countermeasure effectiveness. *Accident Analysis & Prevention, 15(4)*, 287-298.

Hauer, E. (1983). Reflections on methods of statistical inference in research on the effect of safety countermeasures. *Accident Analysis & Prevention 15(4)*, 275-285.

Haus, S., Sherony, R., & Gabler, H. (2019). Estimated benefit of automated emergency braking systems for vehicle–pedestrian crashes in the United States. *Traffic Injury Prevention*, 171-176.

HLDI. (2017). *Predicted availability and fitment of safety features on registered vehicles, Vol. 34, No. 38.* Arlington, VA.: IIHS. Retrieved August 2019, from https://www.iihs.org/media/0a06d47a-84ba-44a0-b661-09f25e81d43d/q78xLA/HLDI%20Research/Bulletins/hldi_bulletin_34.28.pdf

Hoff, P. D. (2009). *A First course in Bayesian statistical methods.* Springer.

Huang, H., & Abdel-Aty, M. (2010). Multilevel data and Bayesian analysis in traffic safety. *Accident Analysis and Prevention 42*, 1556-1565.

Isaksson-Hellman, I., & Lindman, M. (2016). Evaluation of the crash mitigation effect of low-speed automated emergency braking systems based on insurance claims data. *Traffic Inj Prev.*, 42-47.

Jeong, E., & Oh, C. (2017). Evaluating the effectiveness of active vehicle safety systems. *Accident Analysis & Prevention*, 100, 85-96.

Korner, J. (1989). *A Method for Evaluating Occupant Protection by Correlating Accident Data with Laboratory Test Data.* SAE Technical Paper 890747, doi:10.4271/890747.

Kovaceva, J., Bálint, A., Schindler, R., Schneider, A., Stoll, J., Breunig, S., . . . Esquer, A. (2018). *Assessment of the PROSPECT safety systems including socio-economic evaluation.* Deliverable 2.3 in the EU project PROSPECT. Retrieved from https://research.chalmers.se/publication/507588/file/507588_Fulltext.pdf

Krebs, S., Kunert, M., Arbitmann, M., & Othmezouri, G. (2018). *Deliverable 6.2 - Vehicle demonstrators.* Deliverable 6.2 in the EU project PROSPECT. Retrieved from http://www.prospect-project.eu/download/public-files/public_deliverables/PROSPECT-Deliverable-D6.2-Vehicle-demonstrators.pdf

Kreiss, J.-P., Feng, G., Krampe, J., Meyer, M., Niebuhr, T., Pastor, C., & Dobberstein, J. (2015). Extrapolation of GIDAS accident data to Europe. *ESV.* Gothenburg, Sweden: NHTSA.

Kruschke, J. K. (2015). *Doing Bayesian Data Analysis: A Tutorial with R, JAGS and Stan, ed. 2.* Elsevier.

Kuehn, M., Hummel, T., & Bende, J. (2009). Benefit estimation of advanced driver assistance systems for cars derived from real-life accidents. *21st International Technical Conference on the Enhanced Safety of Vehicles ESV*, (p. 18).

Kullgren, A. (2008). Dose-response models and EDR data for assessment of injury risk and effectiveness of safety systems. *IRCOBI.* Bern.

Kusano, K., & Gabler, H. (2012). Safety benefits of forward collision warning, brake assist, and autonomous braking systems in rear-end collisions. *IEEE Transactions on Intelligent Transportation Systems, Volume: 13, Issue: 4*, 1546 - 1555.

Lee, S., Jeong, J., Hong, Y., Shin, J., & Choi, I. (2019). SAFETY EVALUATION OF AUTOMATED VEHICLES THROUGH ACTUAL VEHICLE TESTS IN CUT-IN AND OFFSET CUT-IN SITUATIONS. *ESV.* Eindhoven, NL.

Liddell, T., & Kruschke, J. (2018). Analyzing ordinal data with metric models: What could possibly go wrong? *Journal of Experimental Social Psychology (79)*, 328–348. doi:10.1016/j.jesp.2018.08.009

Lindman, M., Ödblom, A., Bergvall, E., Eidehall, A., Svanberg, B., & Lukaszewicz, T. (2010). Benefit Estimation Model for Pedestrian Auto Brake Functionality. *ESAR.* Hanover, Germany.

McLaughlin, S., Hankey, J., & Dingus, T. (2008). A method for evaluating collision avoidance systems using naturalistic driving data. *Accident Analysis & Prevention*, 40(1), 8-16.

Miaou, S.-P., & Lord, D. (2003). Modeling traffic crash-flow relationships for intersections: dispersion parameter, functional form, and Bayes versus Empirical Bayes methods. *Transportation Research Record 1840*, 31-40.

Mitra, S., & Washington, S. (2007). On the nature of over-dispersion in motor vehicle crash prediction models. *Accident Analysis and Prevention 39*, 459-468.

Morando, A. (2019). *Drivers' Response to Attentional Demand in Automated Driving, PhD Thesis.* Gothenburg, Sweden: Chalmers University of Technology.

Nilsson, J. (2014). *Computational Verification Methods for Automotive Safety Systems.* Gothenburg, Sweden: Institutionen för signaler och system, Mekatronik.

Page, Y., Fahrenkrog, F., Fiorentino, A., Gwehenberger, J., Helmer, T., Lindman, M., & Wimmer, P. (2015). A comprehensive and harmonized method for assessing the effectiveness of advanced driver assistance systems by virtual simulation: THE P.E.A.R.S. initiative. *ESV.* Gothenburg, Sweden.

Persaud, B., Retting, R., Gårder, P., & Lord, D. (2001). Safety Effect of Roundabout Conversions in the United States: Empirical bayes observational before-after study. *Transportation Research Record, 1751(1)*, 1-8.

Petitjean, A., Trosseille, X., Praxl, N., Hynd, D., & Irwin, A. (2012). Injury Risk Curves for the WorldSID 50th Male Dummy. *Stapp Car Crash Journal (56)*, 323-347.

Rosen, E. (2013). Autonomous Emergency Braking for Vulnerable Road Users. *IRCOBI*, (pp. 618-627). Gothenburg, Sweden.

Sander, U. (2018). *Predicting Safety Benefits of Automated Emergency Braking at Intersections.* Gothenburg, Sweden: Chalmers University of Technology. Retrieved from https://research.chalmers.se/publication/504728/file/504728_Fulltext.pdf

Schneider, A. (2016). *Penetration rates in Germany and the US (internal report).* Audi.

Sternlund, S., Strandroth, J., Rizzi, M., Lie, A., & Tingvall, C. (2017). The effectiveness of lane departure warning systems—A reduction in real-world passenger car injury crashes. *Traffic Injury Prevention, 18:2*, 225-229.

van Noort, M., Faber, F., & Bakri, T. (2012). EuroFOT safety impact assessment method and results.

Wang, L., Vogt, T., Dobberstein, J., Bakker, J., Jung, O., Helmer, T., & Kates, R. (2016). Multi-functional open-source simulation platform for development and functional validation of ADAS and automated driving. In R. Isermann (eds.), *Fahrerassistenzsysteme* (pp. 135-148). Proceedings. Springer Vieweg, Wiesbaden.

Wille, J. M., Jungbluth, A., Kohsiek, A., & Zatloukal, M. (2012). rateEFFECT-Entwicklung eines Werkzeugs zur Effizienzbewertung aktiver Sicherheitssysteme. *13 Braunschweiger Symposium AAET.* Braunschweig.

Wimmer, P., During, M., Chajmowicz, H., Granum, F., King, J., Kolk, H., . . . Wagner, M. (2019). Toward harmonizing prospective effectiveness assessment for road safety: Comparing tools in standard test case simulations. *TRAFFIC INJURY PREVENTION*, 139-145.

Wisch, M., Lerner, M., Kovaceva, J., Bálint, A., Gohl, I., Schneider, A., . . . Lindman, M. (2017). Car-to-cyclist crashes in Europe and derivation of use cases as basis for test scenarios of next generation advanced driver assistance systems – results from Prospect. *Enhanced Safety of Vehicles.* Detroit.

Wisch, M., Lerner, M., Schneider, A. J., Attila, G., Kovaceva, J., Bálint, A., & Lindman, M. (2016). *PROSPECT Deliverable 2.1 - Accident Analysis, Naturalistic Driving Studies and Project Implications – Part A: Accident data analyses.* Retrieved from http://www.prospect-project.eu/download/public-files/public_deliverables/PROSPECT-Deliverable-D2.1-Accident-Analysis-NDS-and-Project-Implications.pdf

Xie, S. Q., Dong, N., Wong, S. C., Huang, H., & Xu, P. (2018). Bayesian approach to model pedestrian crashes at signalized intersections with measurement errors in exposure. *Accident Analysis and Prevention 121*, 285-294.

Yanagisawa, M., Swanson, E., Azeredo, P., & Najm, W. G. (2017). *Estimation of potential safety benefits for pedestrian crash avoidance/mitigation systems.* Washington, DC: National Highway Traffic Safety Administration.: Report No. DOT HS 812 400.

Yoganandan, N., Banerjee, A., Hsu, F., Bass, C., Voo, L., Pintar, F., & Gayzik, F. (2016). Deriving injury risk curves using survival analysis from biomechanical experiments. *Journal of Biomechanics*, 3260-3267.

Zhao, Y., Ito, D., & Mizuno, K. (2019). AEB effectiveness evaluation based on car-to-cyclist accident reconstructions using video of drive recorder. *TRAFFIC INJURY PREVENTION, Vol. 20, No.1*, 100-106.

# 8  Acknowledgements

We would like to thank the PROSPECT partners, especially the authors of D2.3, the support of Marcus Wisch (BASt) and Patrick Seiniger (BASt) in the earlier parts of this work, and Carol Flanagan (UMTRI) for fruitful discussions on the Bayesian methodology. We also thank three anonymous reviewers whose valuable comments have substantially improved the paper. The research leading to the results of this work has received funding from the European Community's Eighth Framework Program (Horizon2020) under grant agreement n° 634149. The work has been carried out at SAFER – Vehicle and Traffic Safety Centre at Chalmers, Sweden.

# 9  Appendix

## 9.1  Injury risk functions

An ordered probit regression model was applied to specify the probability of sustaining a certain injury, i.e. slightly injured, seriously injured and fatally injured. The estimator was the collision speed of the car. This model uses the inverse standard normal distribution of the probability as a linear combination of the predictors. The model, with the corresponding coefficients, intercepts and the Akaike Information Criterion (AIC) value, that estimates the injury severity for the cyclist is provided in Table 10. The corresponding model for the pedestrian injuries is provided in Table 11. Note that while only the risk of serious and fatal injuries is considered in this paper, the risk of slight injury is also given for completeness.

**Table 10: Parameter estimates of the ordered probit model for cyclist based on police coded injury severity.**

|  | Estimate | Standard Error | t-value |
|---|---|---|---|
| **Vehicle collision speed** | 0.03197 | 0.002981 | 10.73 |
| **Intercept MAIS1 → MAIS2+** | 1.3679 | 0.0732 | 18.6765 |
| **Intercept MAIS2+ → fatal** | 3.5633 | 0.1949 | 18.2813 |
| **Residual Deviance: 1,426.122** | | | |
| **AIC: 1,432.122** | | | |

**Table 11: Parameter estimates of the ordered probit model for pedestrians based on police coded injury severity.**

|  | Estimate | Standard Error | t-value |
|---|---|---|---|
| **Vehicle collision speed** | 0.03303 | 0.003612 | 9.147 |
| **Intercept MAIS1 → MAIS2+** | 0.8926 | 0.1216 | 7.3398 |
| **Intercept MAIS2+ → fatal** | 3.2316 | 0.1973 | 16.3762 |
| **Residual Deviance: 829.4574** | | | |
| **AIC: 835.4574** | | | |

## 9.2 System algorithms

Table 12: Frequency of total, avoided and mitigated crashes for Algorithm 2 and 3 per use case.

| Use Case | Total | Algorithm 2 | | | | Algorithm 3 | | | |
|---|---|---|---|---|---|---|---|---|---|
| | | Avoided | Mitigated | % Avoided | % Mitigated | Avoided | Mitigated | % Avoided | % Mitigated |
| UC1 | 131 | 111 | 20 | 84.7 | 15.3 | 112 | 19 | 85.5 | 14.5 |
| UC2 | 143 | 51 | 92 | 35.7 | 64.3 | 51 | 92 | 35.7 | 64.3 |
| UC3 | 244 | 101 | 143 | 41.4 | 58.6 | 104 | 140 | 42.6 | 57.4 |
| UC4 | 216 | 125 | 91 | 57.9 | 42.1 | 126 | 90 | 58.3 | 41.7 |
| UC5 | 496 | 391 | 105 | 78.8 | 21.2 | 395 | 101 | 79.6 | 20.4 |
| UC6 | 105 | 81 | 24 | 77.1 | 22.9 | 82 | 23 | 78.1 | 21.9 |
| UC9 | 21 | 15 | 6 | 71.4 | 28.6 | 15 | 6 | 71.4 | 28.6 |
| UC10 | 342 | 117 | 225 | 34.2 | 65.8 | 122 | 220 | 35.7 | 64.3 |
| UC11 | 216 | 34 | 182 | 15.7 | 84.3 | 36 | 180 | 16.7 | 83.3 |
| UC12 | 20 | 5 | 15 | 25.0 | 75.0 | 5 | 15 | 25.0 | 75.0 |

Table 13: Frequency of avoided, mitigated and total number of crashes for Algorithm 4 (steering and braking).

| Use Case | Total | Avoided | Mitigated | % Avoided | % Mitigated |
|---|---|---|---|---|---|
| UC9 | 21 | 19 | 2 | 90.5 | 9.5 |
| UC12 | 20 | 16 | 4 | 80.0 | 20.0 |

Table 14: Posterior models for the probability of crash avoidance for Algorithm 2 and 3.

| | Algorithm 2 | | Algorithm 3 | |
|---|---|---|---|---|
| | Coefficient of car initial speed $\beta_1$ | Intercept $\beta_0$ | Coefficient of car initial speed $\beta_1$ | Intercept $\beta_0$ |
| UC1 | -0.205 | 5.751 | -0.205 | 5.774 |
| UC2 | 0.033 | -0.957 | 0.033 | -0.957 |
| UC3 | -0.078 | 1.843 | -0.078 | 1.903 |
| UC4 | -0.049 | 1.997 | -0.052 | 2.121 |
| UC5 | -0.041 | 1.801 | -0.042 | 1.869 |
| UC6 | 0.018 | 1.260 | 0.013 | 1.371 |
| UC9 | -0.915 | 50.803 | -0.915 | 50.803 |
| UC10 | -0.086 | 1.963 | -0.088 | 2.068 |
| UC11 | -0.044 | -0.454 | -0.049 | -0.236 |
| UC12 | -0.370 | 14.146 | -0.370 | 14.146 |

Table 15: Maximum potential reduction of cyclist injuries for one year by different algorithms for UC9. The estimate is given together with the lower and higher bound interval (in brackets).

| | UC9 | | | |
|---|---|---|---|---|
| | Algorithm 1 – braking | Algorithm 2 – braking | Algorithm 3 – braking | Algorithm 4 – steering and braking |
| Fatal | 93 (85-96) | 87 (78-91) | 87 (78-91) | 102 |
| Serious | 258 (219-263) | 251 (203-255) | 251 (205-254) | 294 |

Table 16: Maximum potential reduction of pedestrian injuries for one year by different algorithms for UC12. The estimate is given together with the lower and higher bound interval (in brackets).

|  | UC12 | | | |
|---|---|---|---|---|
|  | Algorithm 1 – braking | Algorithm 2 – braking | Algorithm 3 – braking | Algorithm 4 – steering and braking |
| **Fatal** | 70 (65-76) | 64 (59-74) | 65 (59-74) | 78 (59-82) |
| **Serious** | 287 (190-400) | 266 (152-395) | 269 (154-396) | 487 (364-515) |

## 9.3 Decision trees

The result of the decision tree method is a structure that represents the GIDAS database in a tree format, see Figure 5 for cyclist injuries and Figure 6 for pedestrian injuries. Each node has a node number, and the class of that node (i.e. fatal, serious or slight). The three values under the class show the number of cases to each class for that node.

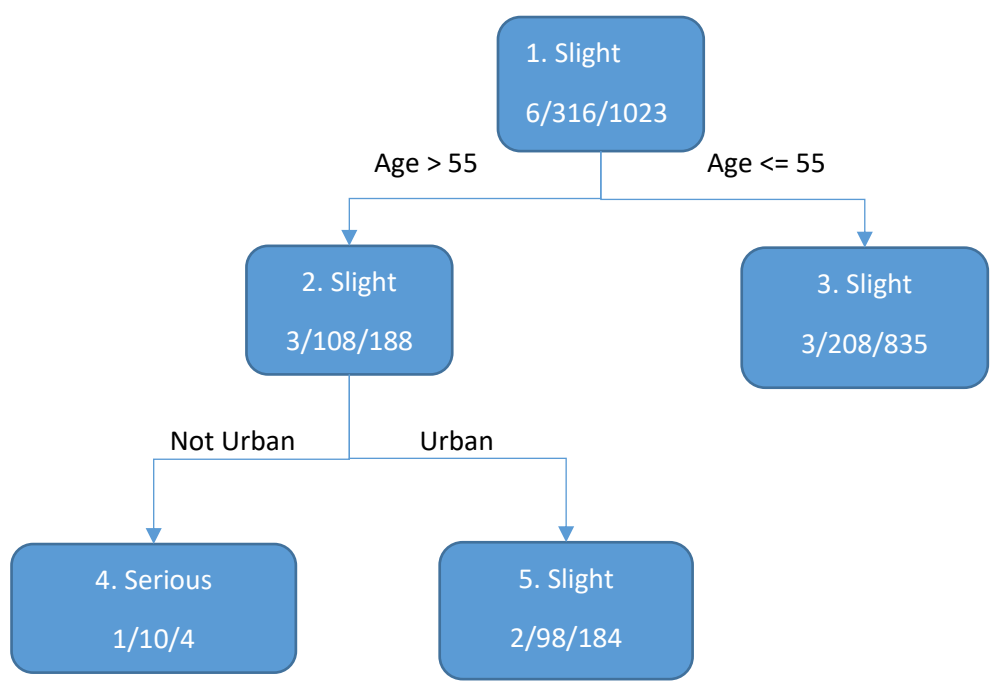

Figure 5: Classification tree for cyclist injuries from GIDAS. In each box the first row shows the node number and the predicted injury class for that node; the second row shows the number of cases that are classified as fatal/serious/slight.

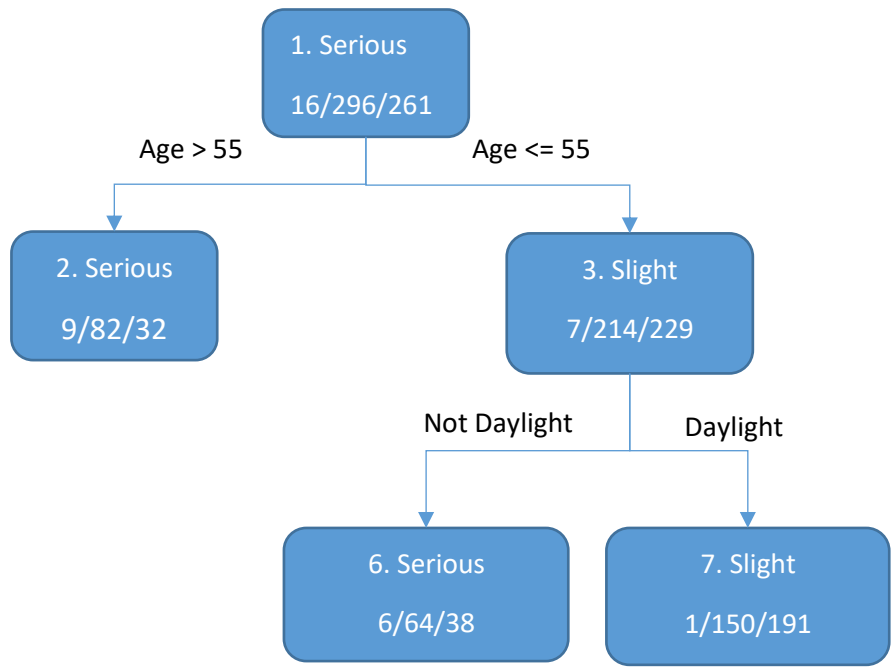

**Figure 6: Classification tree for pedestrian injuries from GIDAS. In each box the first row shows the node number and the predicted injury class for that node; the second row shows the number of cases that are classified as fatal/serious/slight.**